%% file: eprint.tex
\documentclass[10pt]{article}
\usepackage{graphicx}
\pdfoutput=1

\def\Title#1{\begin{center} {\Large #1 } \end{center}}
\def\Author#1{\begin{center}{ \sc #1} \end{center}}
\def\Address#1{\begin{center}{ \it #1} \end{center}}

\newcommand\pubblock{\rightline{\begin{tabular}{l} Proceedings of the Second Annual LHCP\\ \pubnumber\\
         \pubdate  \end{tabular}}}

\newenvironment{Abstract}{\begin{quotation} \begin{center}
             \large ABSTRACT \end{center}\bigskip 
      \begin{center}\begin{large}}{\end{large}\end{center} \end{quotation}}

\newenvironment{Presented}{\begin{quotation} \begin{center} 
             PRESENTED AT\end{center}\bigskip 
      \begin{center}\begin{large}}{\end{large}\end{center} \end{quotation}}

\def\Acknowledgements{\bigskip  \bigskip \begin{center} \begin{large}
             \bf ACKNOWLEDGEMENTS \end{large}\end{center}}

\input econfmacros.tex

\usepackage{color}
\usepackage{subfigure}

\newcommand\pubnumber{ ATL-PHYS-PROC-2014-108 }

\newcommand\pubdate{\today}

\def\affiliation{
On behalf of the ATLAS, CDF and CMS Experiments, \\
Department of Physics and Astronomy, High Energy Physics \\
Uppsala University, SE-751 20 Uppsala, Sweden}

\begin{document}

\large
\begin{titlepage}
\pubblock

\vfill
\Title{ Physics with boosted top quarks  }
\vfill

\Author{ Elin Bergeaas Kuutmann  }
\Address{\affiliation}
\vfill
\begin{Abstract}

The production at the LHC of \emph{boosted top quarks} (top quarks with a transverse momentum that greatly exceeds their rest mass) is a promising process to search for phenomena beyond the Standard Model. 
In this contribution several examples are discussed of new techniques to reconstruct and identify (\emph{tag}) the collimated decay topology of the boosted hadronic decays of top quarks. 
Boosted top reconstruction techniques have been utilized in searches for new physical phenomena. 
An overview is given of searches by ATLAS, CDF and CMS for heavy new particles decaying into a top and an anti-top quark, vector-like quarks and supersymmetric partners to the top quark.
\end{Abstract}
\vfill

\begin{Presented}
The Second Annual Conference\\
 on Large Hadron Collider Physics \\
Columbia University, New York, U.S.A \\ 
June 2-7, 2014
\end{Presented}
\vfill
\end{titlepage}
\def\thefootnote{\fnsymbol{footnote}}
\setcounter{footnote}{0}
%

\normalsize 


\section{Introduction}

If the Standard Model (SM) is the final theory of elementary particle physics, the observed mass of the Higgs boson must be fine-tuned. 
There are several proposed extensions to the SM, that offer natural explanations to the Higgs boson mass, 
e.g.~supersymmetry, new strong dynamics (Technicolor and other models), models with extra dimensions (such as Randall--Sundrum), the existence of a fourth generation of quarks or vector-like quarks, and several other scenarios (see, e.g.,~\cite{ATLAS:2013kha,Chatrchyan:2013lca} and references therein). 
All these theories tend to come with top partners or particles decaying to top quarks. 
However, no such particles have yet been observed, which means that if they exist, they are likely heavy. 

If the decay of a heavy exotic particle contains one or more top quarks, these top quarks will be naturally boosted from the parent mass. 
For a high transverse momentum of the top quark, the three decay products of the top quark ($\ell \nu b$ or $q\bar{q}'b$) will be collimated. 
The typical radius of the containment cone\footnote{Hadron collider detectors typically have a cylindrical coordinate system with the $z$-axis along the beampipe. $\phi$ is the azimuthal angle and $\theta$ the polar angle. The pseudorapidity $\eta$ is defined as $\eta = - \ln \tan (\theta /2)$. Angular distances are given as $\dR = \sqrt{\Delta \eta ^2 + \Delta \phi ^2}$.} 
 of the decay is $R = 2m/\pt $. 
If $\pttop > 350$~\GeV, the top decay will be contained in one single jet with radius parameter $R=1.0$, and an attempt to reconstruct $t\rightarrow q\bar{q}'b$ as three separate smaller jets (\emph{resolved reconstruction}) will fail. 
Special techniques are needed to recover these events. 

In the following, recent analysis results with boosted top quarks from the  LHC~\cite{Evans:2008zzb} experiments ATLAS~\cite{Aad:2008zzm} and CMS~\cite{Chatrchyan:2008aa}  will be shown, as well as a physics analysis from CDF. 
A good overview of existing techniques for boosted top quark reconstruction can be obtained from the BOOST workshop proceedings \cite{Abdesselam:2010pt,Altheimer:2012mn,Altheimer:2013yza}. 

\section{Boosted top quarks at the Tevatron}

The first attempt to measure the production cross section of boosted top quarks was made by the CDF experiment~\cite{CDF:10234} using 6~\ifb\ of \ppbar\ collision data at $\sqrts = 1.96$~\TeV. 
\begin{figure}[htb]
\centering
\subfigure[Simulated mass of the leading jet, \mjetone. \label{F:mjet1_qcd_ttbar_SMET_gt_4_Midpoint_c1}]{
\includegraphics[width=0.466\textwidth]{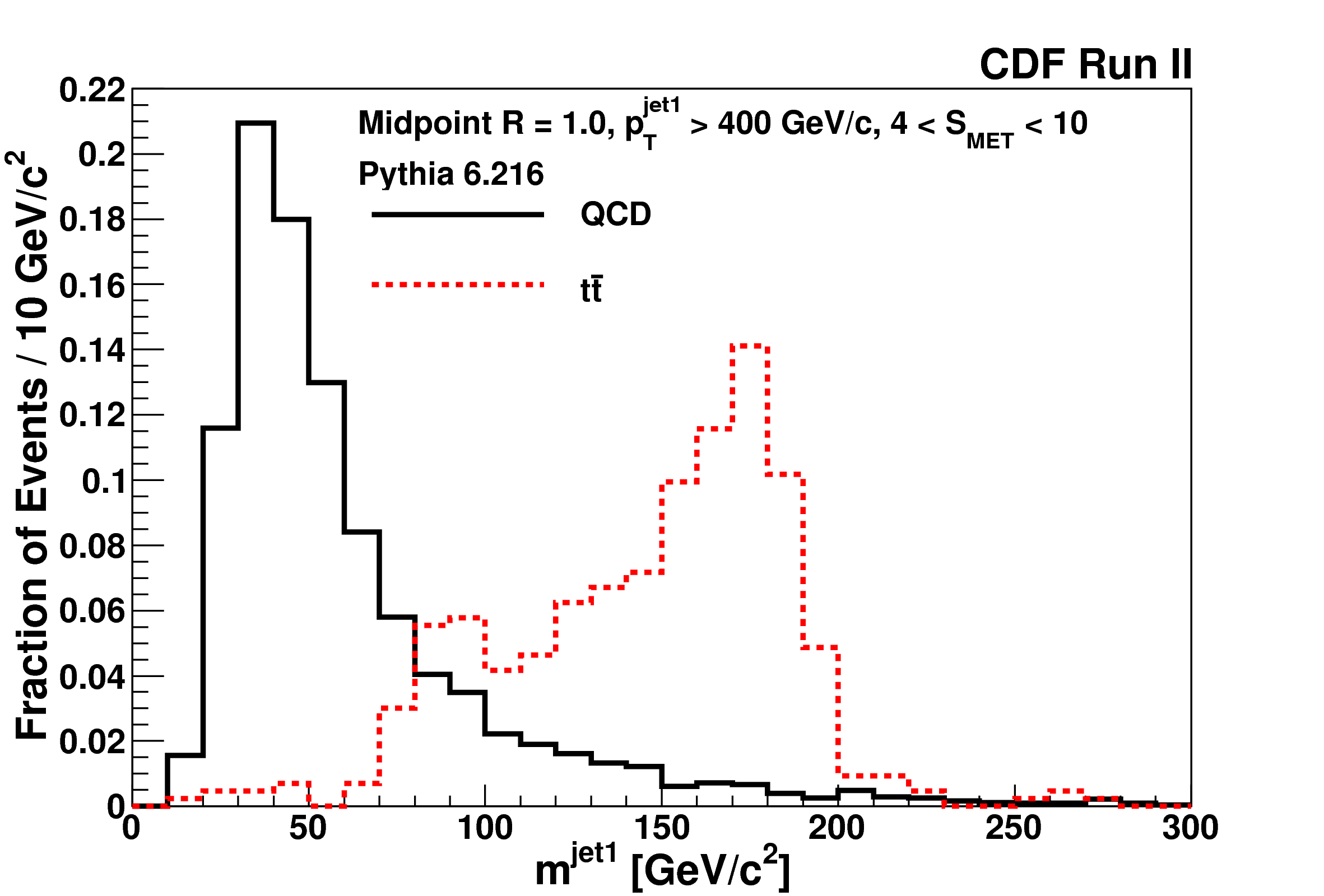} }
\hspace{2em}
\subfigure[\mjettwo\ vs.~\mjetone, data. \label{F:mjet2_vs_mjet1_fullHadronic_data_Midpoint}]{
\includegraphics[width=0.390\textwidth]{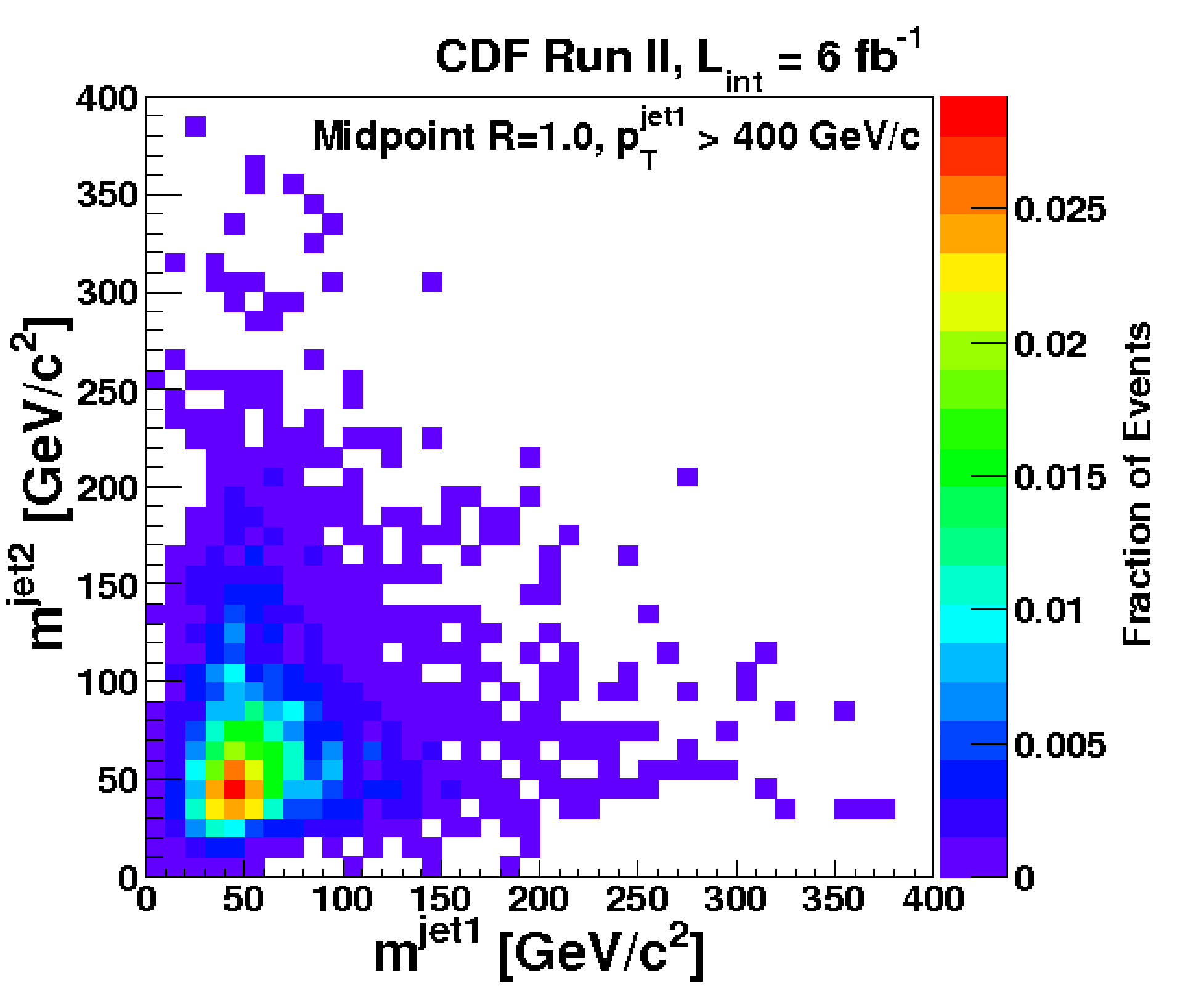} }
\caption{Jet mass: (a) mass of the leading jet, \mjetone, from simulations of QCD and \ttbar\ production and (b) the data distribution of the subleading jet mass \mjettwo\ vs.~the leading jet mass \mjetone. 
Figures from Ref.~\cite{CDF:10234}. }
\label{fig:cdf10234}
\end{figure}
In this study, boosted top quarks are identified through the mass of jets with radius parameter $R=1.0$. 
This mass is expected to coincide with the top quark mass ($\sim 173$~\GeV) for \ttbar\ production, while being lower for multi-jets, as illustrated in Fig.~\ref{F:mjet1_qcd_ttbar_SMET_gt_4_Midpoint_c1}.
The event selection requires either 2 massive jets ($130<\mjet<210$~\GeV), or 1 massive jet and the presence of \MET\ (\emph{missing transverse energy}\footnote{\MET is the component needed to bring the vector sum of the transverse momenta of all reconstructed objects to 0. In \ttbarljets\ events, this corresponds to the undetected energy of the neutrino. }). 
The mass distribution in data in the all-hadronic channel is shown in Fig.~\ref{F:mjet2_vs_mjet1_fullHadronic_data_Midpoint}. 
A total of 58 events are observed in both channels, with an expectation of $44 \pm 8 \mathrm{(stat)} \pm 13 \mathrm{(syst)}$, of which only five events are expected to originate from \ttbar\ production rather than QCD. 
An upper 95\% CL limit on \ttbar\ production for $\pttop>400$~\GeV\ of $\sigma < 40$~fb is set. 

\section{Boosted top quarks at the LHC}

The higher production cross section of \ttbar\ at the LHC compared with the Tevatron, makes studies of boosted top quarks at the LHC easier. 
For an overview of existing techniques for boosted top reconstruction, the performance notes of CMS~\cite{CMS:2014fya} and ATLAS~\cite{ATLAS:2013qia} are recommended. 

Both ATLAS and CMS have made interesting performance studies of various techniques, such as shower deconstruction~\cite{ATLAS:showerdeconstr}, jet grooming~\cite{ATLAS:2013qia,Chatrchyan:2013vbb,Peruzzi:2014here} 
and $N$-subjettiness~\cite{CMS:2014fya,ATLAS:2013qia}, which will not be further discussed here. 
$b$-tagging at high \pt\ is an important aspect of boosted top studies. 
Details can be found in Ref.~\cite{CMS:2013vea}. 

The existing physics analyses using boosted top quarks at the LHC are searches for new heavy particles decaying into top quarks (\emph{resonances}), supersymmetric top squarks and vector-like quarks, to be discussed in the following.

\subsection{Searches for top quark resonances}

A new particle decaying to \ttbar, $X \rightarrow \ttbar$, would be a clear indication of physics beyond the Standard Model. 
Two benchmark theories with $X \rightarrow \ttbar$ have been used at the LHC: a leptophobic topcolor \Zprime\ and a Kaluza--Klein gluon (\gKK). 
ATLAS and CMS use slightly different parameters in the \gKK\ model, hence only \Zprime\ results will be discussed here for comparison. 
Details on the models can be found in Refs.~\cite{ATLAS:2013kha} and~\cite{Chatrchyan:2013lca}. 

\subsubsection{CMS \ttbar\ resonances}

The CMS experiment has searched for \ttbar\ resonances using the full 2012 data set (20~\ifb\ of \sqrts = 8~\TeV\ collisions), combining the \ttbarljets\ (resolved and boosted) and \ttbaralljets\ (boosted only) channels~\cite{Chatrchyan:2013lca}. 
\begin{figure}[htb]
\centering
\subfigure[The \mttbar\ spectrum, $\ell + \jets$, 0 $b$-tag channel. \label{F:CMS-B2G-13-001_Fig1a}]{
\includegraphics[width=0.37\textwidth]{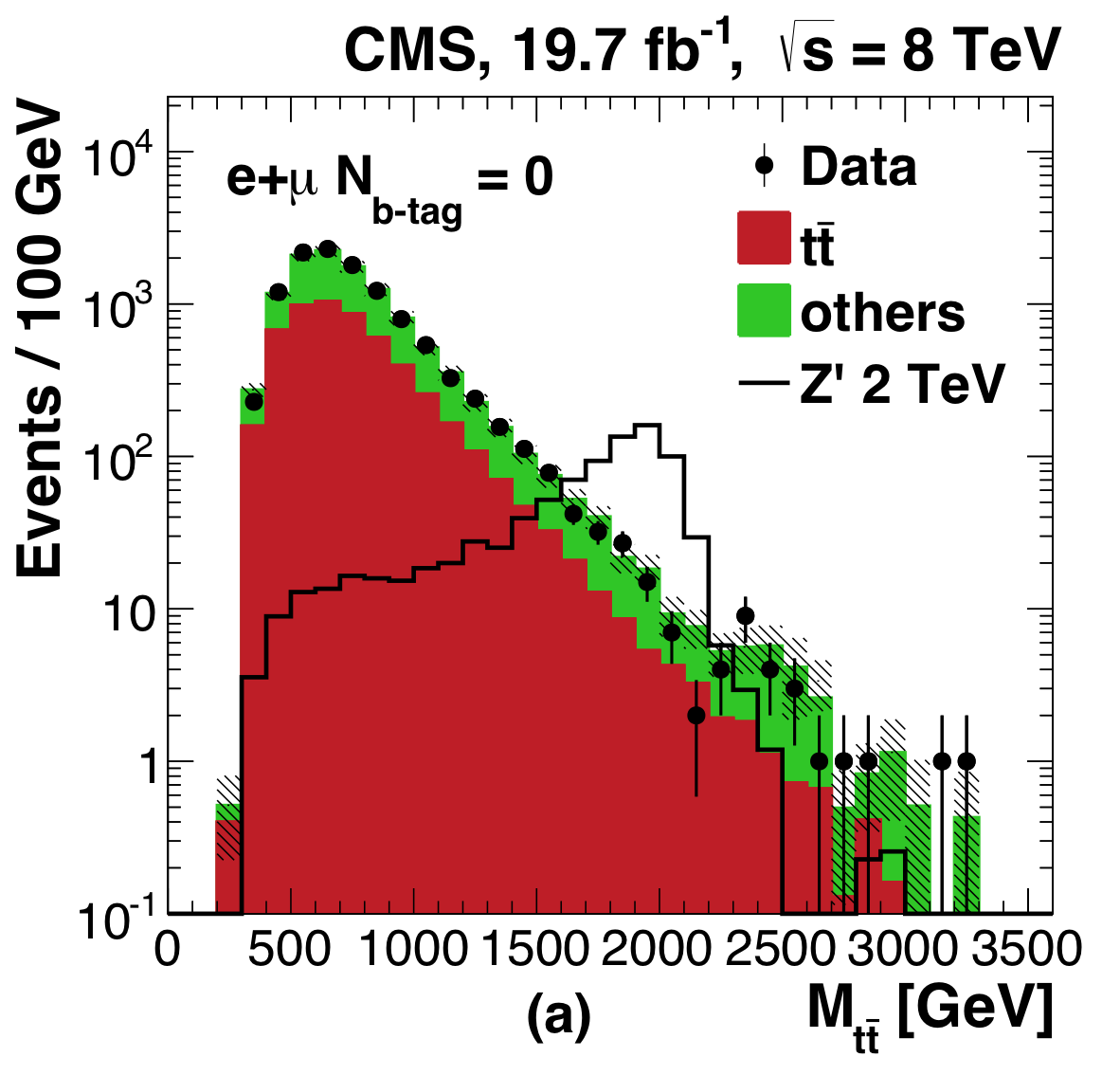} }
\subfigure[Combined upper cross section limits. \label{F:CMS-B2G-13-001_Fig2}]{
\includegraphics[width=0.53\textwidth]{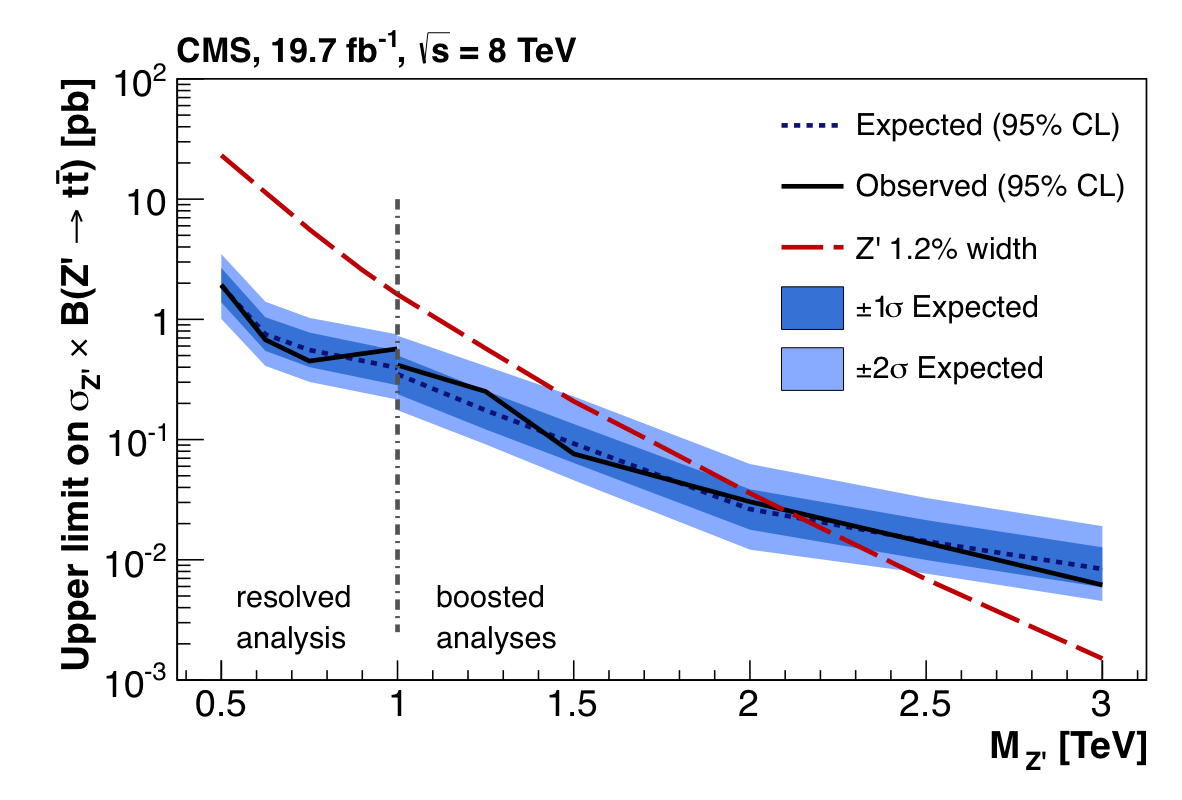} 
}
\caption{(a) Example of a reconstructed \mttbar\ spectrum, here the boosted $\ell + \jets$, 0 $b$-tag channel. 
(b) The upper cross section limit on $\Zprime \rightarrow \ttbar$ production, using all channels. 
Figures from Ref.~\cite{Chatrchyan:2013lca}.
\label{fig:CMS-B2G-13-001_2}}
\end{figure}

For the resolved $\ell+\jets$ channel, the presence of exactly one isolated electron or muon is required, as well as \MET\ and $\geq 4$ anti-\kt\ jets~\cite{Cacciari:2008gp} with $R=0.5$. 
At least one of these jets must be $b$-tagged. 
In the boosted $\ell+\jets$ channel, the charged leptons are not required to be isolated. 
\MET\ must be present, as well as at least two high-\pt\ anti-\kt\ $R=0.5$ jets. 
There are two $b$-tag categories: no tag, or at least one of the jets tagged. 
The dominant non-\ttbar\ background after these cuts is $W$+jets production. 
Figure~\ref{F:CMS-B2G-13-001_Fig1a} shows the \mttbar\ spectrum for the boosted 0 $b$-tag channel. 

In the boosted all-hadronic channel, two jets are required, that have passed the \emph{CMS top-tagger}~\cite{CMS:2014fya}. 
This tagger is based on the \emph{Johns-Hopkins top-tagger}~\cite{Kaplan:2080ie} and starts from Cambridge--Aachen (C--A) jets~\cite{Dokshitzer:1997in} with $R=0.8$. 
The jets are declustered, looking for substructure compatible with the three-pronged \tqqb\ decay. 
The two CMS-tagged top candidate jets must be central in the detector, fulfill $\pt>400$~\GeV\ and be separated by $\Delta \phi(\jet,\jet) > \pi/2$. 
The dominant non-\ttbar\ background after this selection is non-top multi-jets. 

The combination of the boosted and the resolved channels is done through a mass cut-off: for \mbox{$m_{\Zprime}>1$~\TeV}, the boosted selection is used and below this the resolved selection, as can be seen in Fig.~\ref{F:CMS-B2G-13-001_Fig2}. 

The mass exclusion for the leptophobic \Zprime\ is $m<2.1$~\TeV\ (observed and expected).

\subsubsection{ATLAS \ttbarljets\ resonances}

The ATLAS experiment has searched for resonances decaying to \ttbarljets\ using 14~\ifb\ of 8~\TeV\ $pp$ collision data~\cite{ATLAS:2013kha}.
\begin{figure}[htb]
\centering
\subfigure[The reconstructed \mttbar\ spectrum. \label{F:ATLAS-CONF-2013_fig_09}]{
\includegraphics[width=0.46\textwidth]{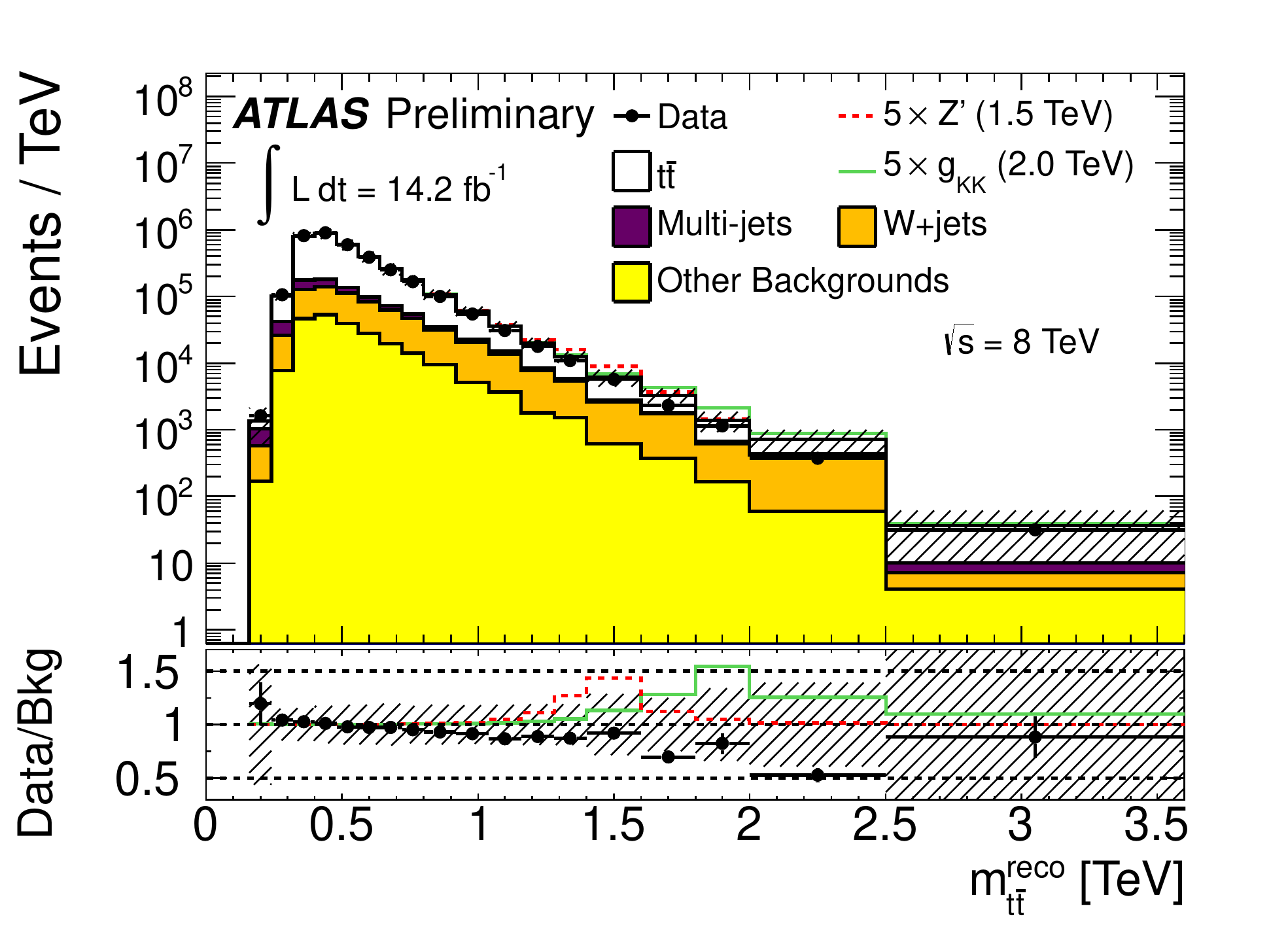} }
\subfigure[Upper cross section limit on $\Zprime \rightarrow \ttbar$ production. \label{F:ATLAS-CONF-2013_fig_10a}]{
\includegraphics[width=0.44\textwidth]{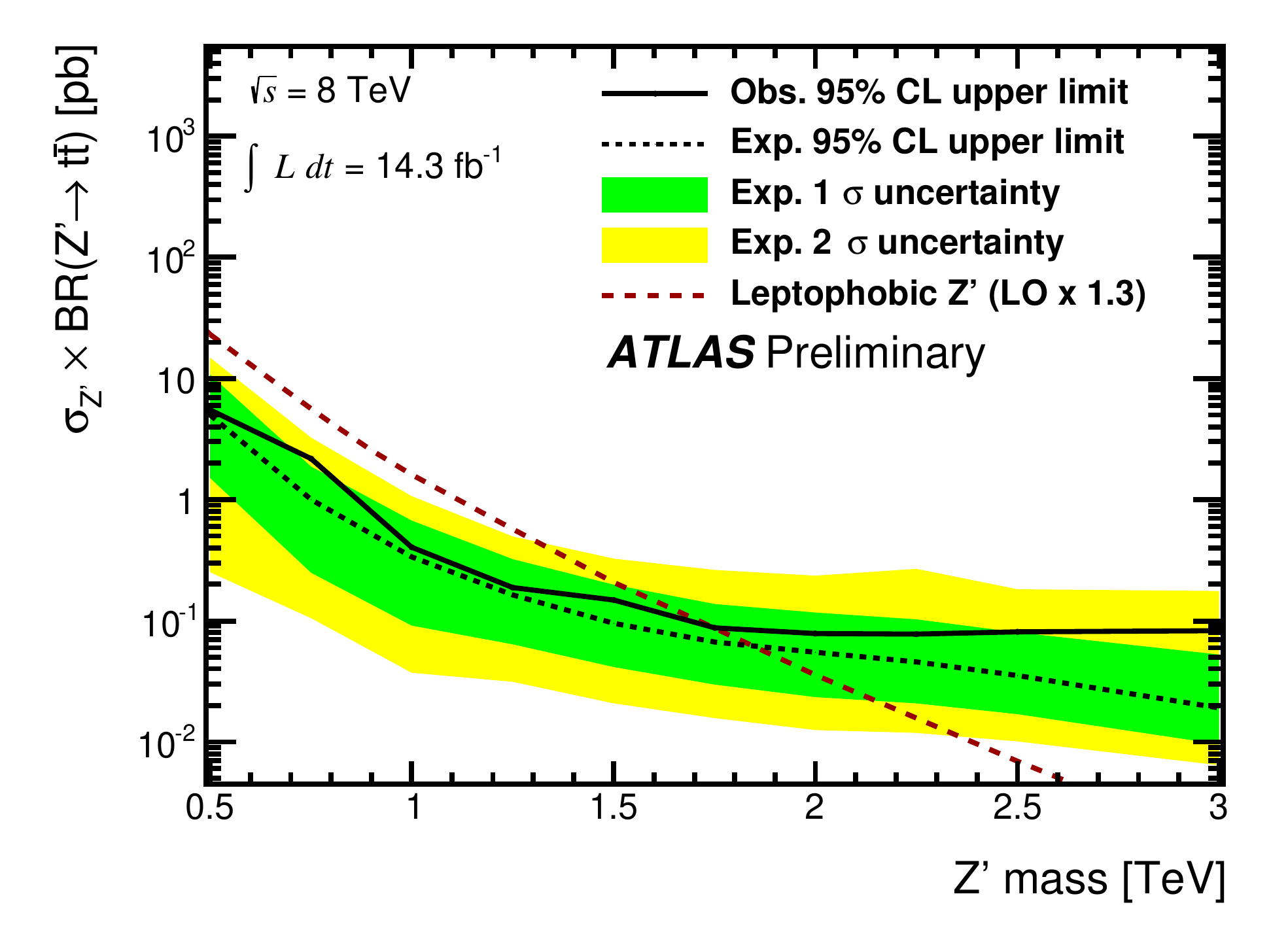} }
\caption{(a) The invariant \ttbar\ mass spectrum with the hypothetical resonances visible as peaks and (b) the upper cross section limit on $\Zprime \rightarrow \ttbar$ production as a function of \Zprime\ mass. Figures from Ref.~\cite{ATLAS:2013kha}. 
\label{fig:ATLAS-CONF-2013}}
\end{figure}
Events are selected by requiring exactly one electron or muon, \MET\ and at least one $b$-tagged anti-\kt\ $R=0.4$ jet anywhere in the event. 
The charged leptons are \emph{mini-isolated}~\cite{Rehermann:2010vq} using a shrinking-cone isolation technique that maintains a stable selection efficiency also at high top quark \pt, when the leptons and the $b$-quark in the \tlnub\ decay are collimated. 
Reconstruction techniques adapted for both boosted top quarks and the more well-known low-energy resolved case are used. 
A boosted hadronic top object is defined as trimmed~\cite{ATLAS:2013qia} anti-\kt\ $R=1.0$ jet that fulfills $\mjet > 100$~\GeV, $\ptjet > 300$~\GeV, \mbox{$\Delta \phi(\ell, \jet)>2.3$} and $\sqrt{d_{12}} > 40$~\GeV, where $d_{12}$ is the first splitting scale of the \kt\ clustering history~\cite{Catani:1993hr,Ellis:1993tq,Cacciari:2011ma}. 
If the event contains a boosted hadronic top object, it ends up in the ``boosted'' category. 
If not, it is tested for a resolved \ttbar\ selection with several anti-\kt\ $R=0.4$ jets, thus creating a seamless transition between reconstructions optimized for low and high \mttbar, shown in Fig.~\ref{F:ATLAS-CONF-2013_fig_09}. 
The mass exclusion of the \Zprime\ model is $0.5 <m< 1.8$~\TeV\ (observed) and $0.5 <m< 1.9$~\TeV (expected).

\subsubsection{ATLAS \ttbar\ resonances, all-hadronic decays}

Searches for \ttbaralljets\ resonances have been done by the ATLAS experiment using the full 2011 LHC dataset~\cite{Aad:2012raa} (5~\ifb\ of 7~\TeV\ $pp$ collisions). 
In this analysis, two different top-tagging techniques were used, \emph{HEPTopTagger}~\cite{Plehn:2009rk,Plehn:2010st} and the \emph{Top Template Tagger}~\cite{Almeida:2011aa}. 

The HEPTopTagger algorithm~\cite{Plehn:2009rk,Plehn:2010st} starts from C-A jets with $R=1.5$ and $\pt>200$~\GeV. 
The clustering history is investigated until light sub-jets are found. 
After several steps of grooming and reclustering, three sub-jets have been created, that are declared a top candidate, if the jet mass ratios are compatible with a $t \rightarrow W b \rightarrow q\bar{q'}b$ decay. 
The algorithm has a stable performance under increasing pile-up. 
To identify the \ttbaralljets\ events, 2 HEPTopTagged jets are required, as well as the existence of 2 $b$-tagged anti-\kt\ $R=0.4$ jets embedded in the large-$R$ jets. 
In Fig.~\ref{F:ATL-TOPQ-2012-15_fig_13a}, the resulting \mttbar\ spectrum is shown. The multijet non-top background is well suppressed. 
\begin{figure}[htb]
\centering
\subfigure[\mttbar\ spectrum, HEPTopTagger method. \label{F:ATL-TOPQ-2012-15_fig_13a}]{
\includegraphics[width=0.37\textwidth]{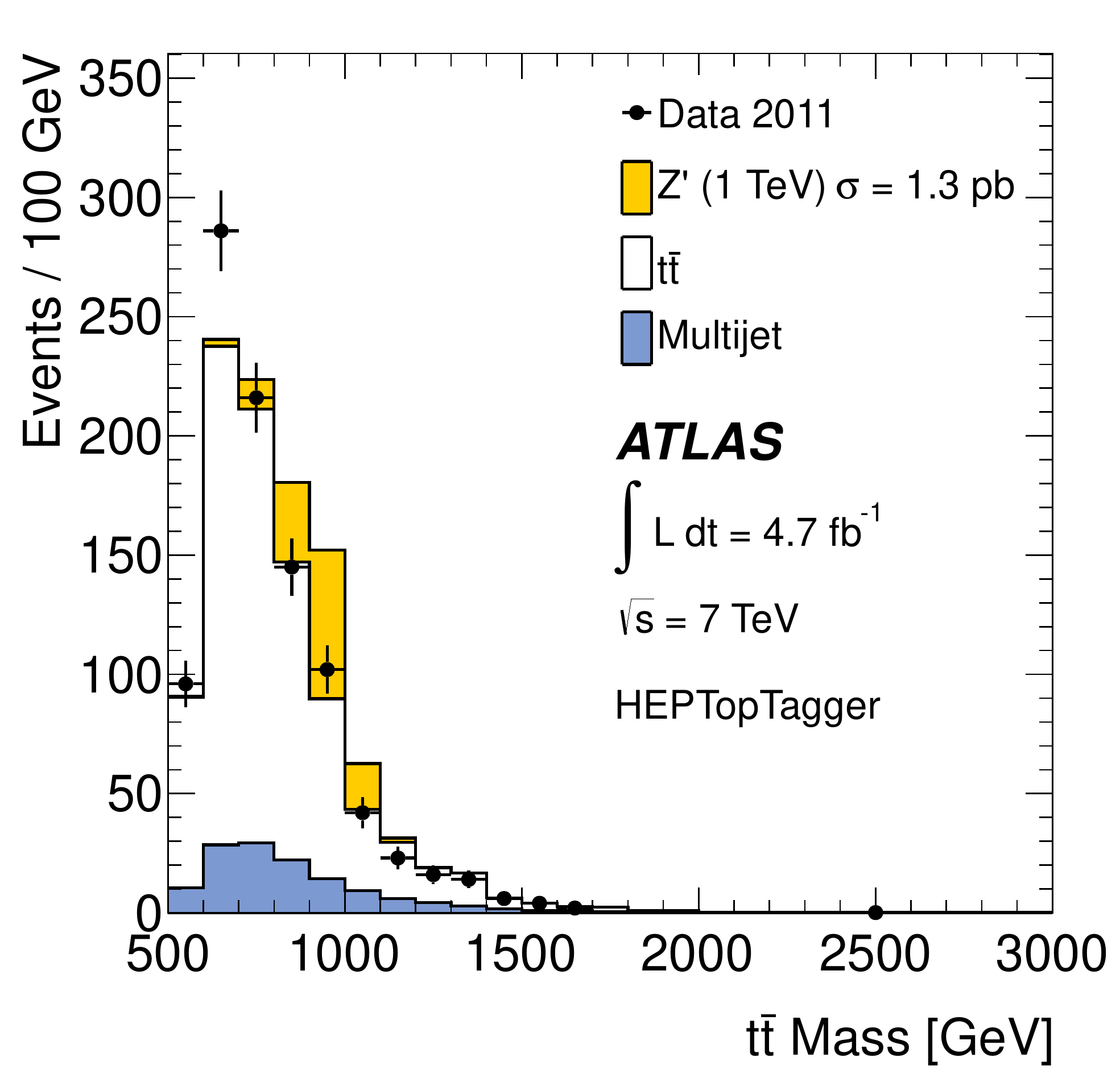} }
\hspace{5em}
\subfigure[\mttbar\ spectrum, Top Template Tagger method.  \label{F:ATL-TOPQ-2012-15_fig_13b}]{
\includegraphics[width=0.37\textwidth]{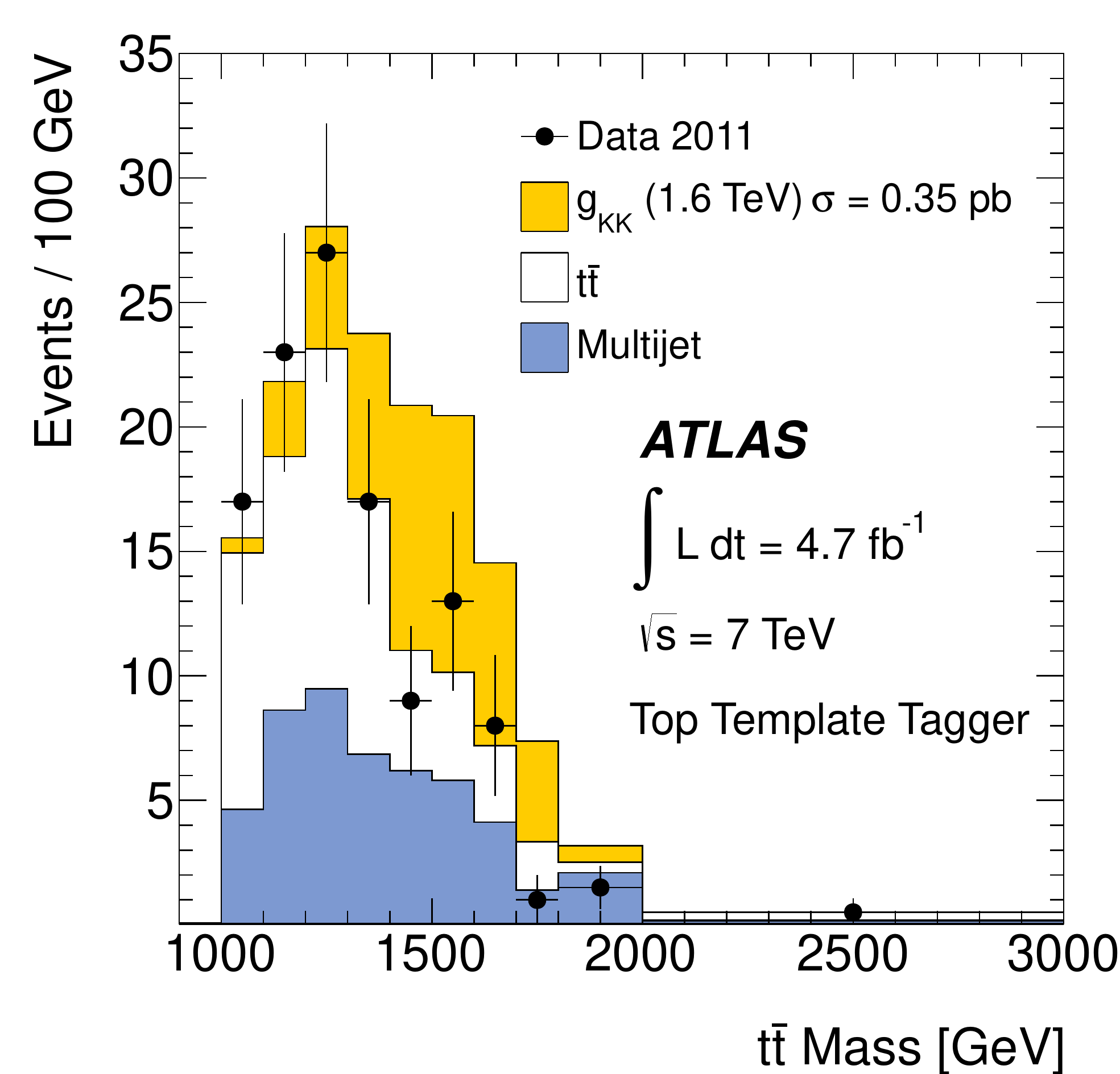} }
\caption{Invariant \ttbaralljets\ mass spectra, reconstructed with (a) the HEPTopTagger algorithm and (b) the Top Template Tagger method. 
The $x$- and $y$-axes are different in the plots, as well as the signal samples compared, the leptophobic \Zprime\ in (a) and the Kaluza--Klein gluon in (b). 
Figures from Ref.~\cite{Aad:2012raa}. 
\label{fig:ATL-TOPQ-2012-15}}
\end{figure}

The Top Template Tagger~\cite{Almeida:2011aa} works with the cluster structure in anti-\kt\ $R=1.0$ jets. 
The energy of the topological clusters in the jets is compared with the parton kinematics in simulated templates of $\ttbar\rightarrow q\bar{q}'b q\bar{q}'b$  decays, by defining an overlap function, $OV_3$, that is close to 1 for a perfect match and close to 0 for a typical multi-jet event without three-prong structure.  

The event selection requires at least two anti-\kt\ $R=1.0$ jets with $\pt>500$~\GeV\ and $\pt>450$~\GeV, respectively, and $|\eta|<2.0$. 
The overlap function must be $OV_3 > 0.7$ for both jets.  
The jet masses of the two jets must coincide with the top quark mass within $\pm 50$~\GeV, and the event must contain two $b$-tagged anti-\kt\ $R=0.4$ jets embedded in the large-$R$ jets. 
The invariant \ttbar\ mass spectrum is shown in Fig.~\ref{F:ATL-TOPQ-2012-15_fig_13b}. 

Comparing the two tagger methods, HEPTopTagger works best for moderately boosted top quarks, while the Top Template Tagger has a slightly better performance for high invariant \ttbar\ masses.

\subsubsection{Summary of the \ttbar\ resonances studies at the LHC}
All \ttbar\ resonance searches at the Tevatron and the LHC, boosted and resolved, are summarized in Ref.~\cite{BOOST2012_Fig8_update}. 
The LHC experiments have increased the scope of resonance searches compared with the Tevatron, probing mass regions well above 1~\TeV. 
It has been demonstrated at the LHC that the all-hadronic final states of \ttbar\ decays are also accessible, with almost the same sensitivity as the \ljets\ channel, due to the development of jet substructure techniques and top-tagging.

\subsection{Searches for top partners}

Many proposed extensions to the SM contain top partners, hypothetical particles that resemble the top quark in some way, such as the supersymmetric top squark or the vector-like top quark, \Tprime. 
In many scenarios, the top partners are heavy and decay into at least one top quark, leading to potentially boosted top quarks in the final state. 
 
A HEPTopTagger-like tagger has been used in a CMS analysis~\cite{CMS:2013nia} looking for pair-produced top squarks decaying into a weakly interacting massive particle and a top quark. 
An ATLAS analysis, also searching for pair-produced top squarks~\cite{Aad:2014bva}, instead reclusters anti-\kt\ $R=0.4$ jets with radii $R=0.8$ and $R=1.2$, using the masses of the large-$R$ jets to find top quark decays and $W$ bosons. More details are given in Ref.~\cite{Wanotayaroj:2014here}. 

A search for pair-produced vector-like quarks~\cite{Aguilar-Saavedra:2013qpa} of exotic charge $\pm 5e/3$ has been done by the CMS experiment~\cite{Chatrchyan:2013wfa}. 
The production and decay $\Tfivethree\Tfivethree \rightarrow W^+ t W^- t$ is illustrated in Fig.~\ref{F:CMS-PAS-B2G-12-012_T53Decay}. 
\begin{figure}[htb]
\centering
\subfigure[ \label{F:CMS-PAS-B2G-12-012_T53Decay}]{
\includegraphics[width=0.37\textwidth]{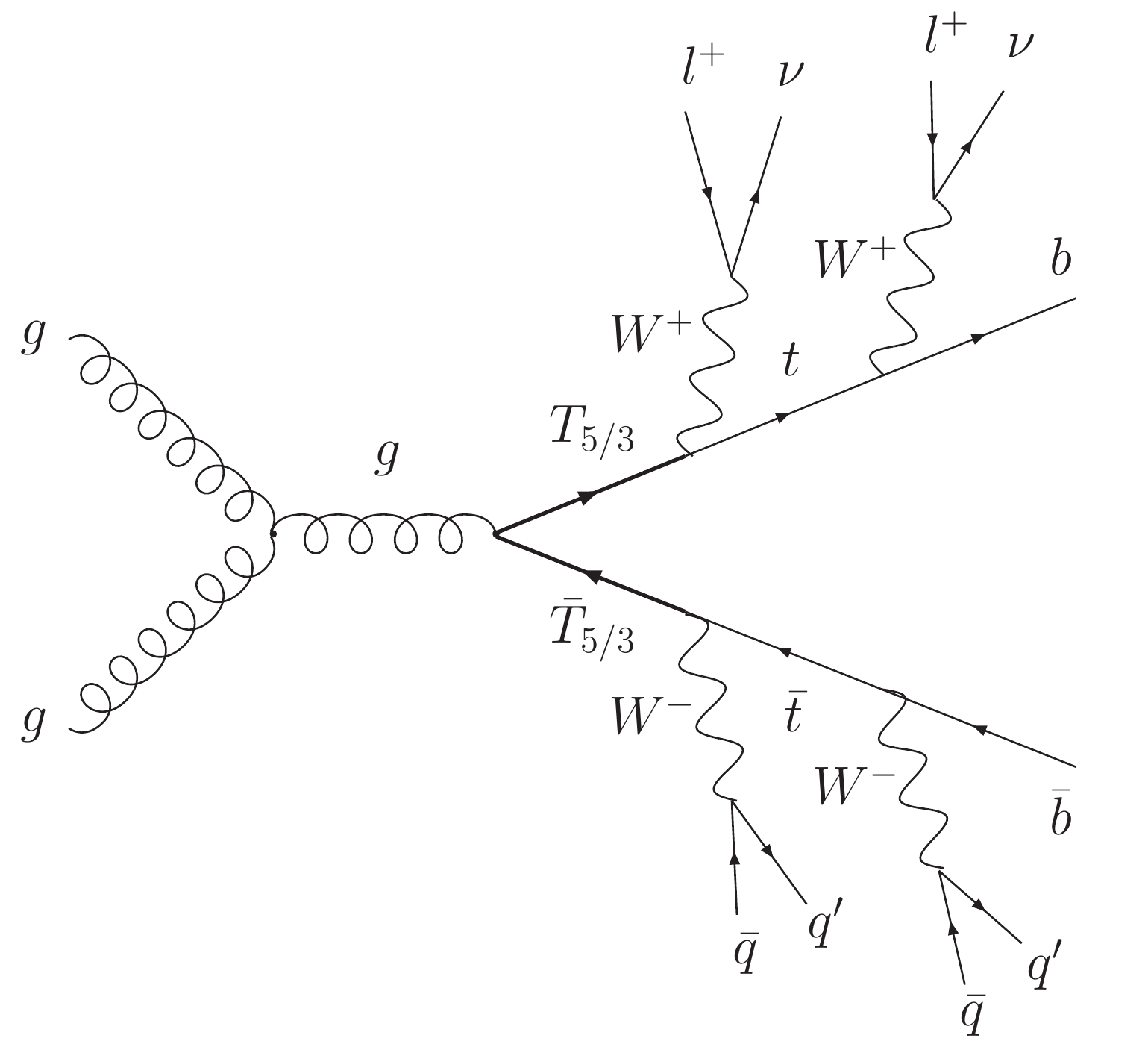} 
\vspace{2em}}
\hspace{7em}
\subfigure[\label{F:CMS-PAS-B2G-12-012_Limit}]{
\includegraphics[width=0.37\textwidth]{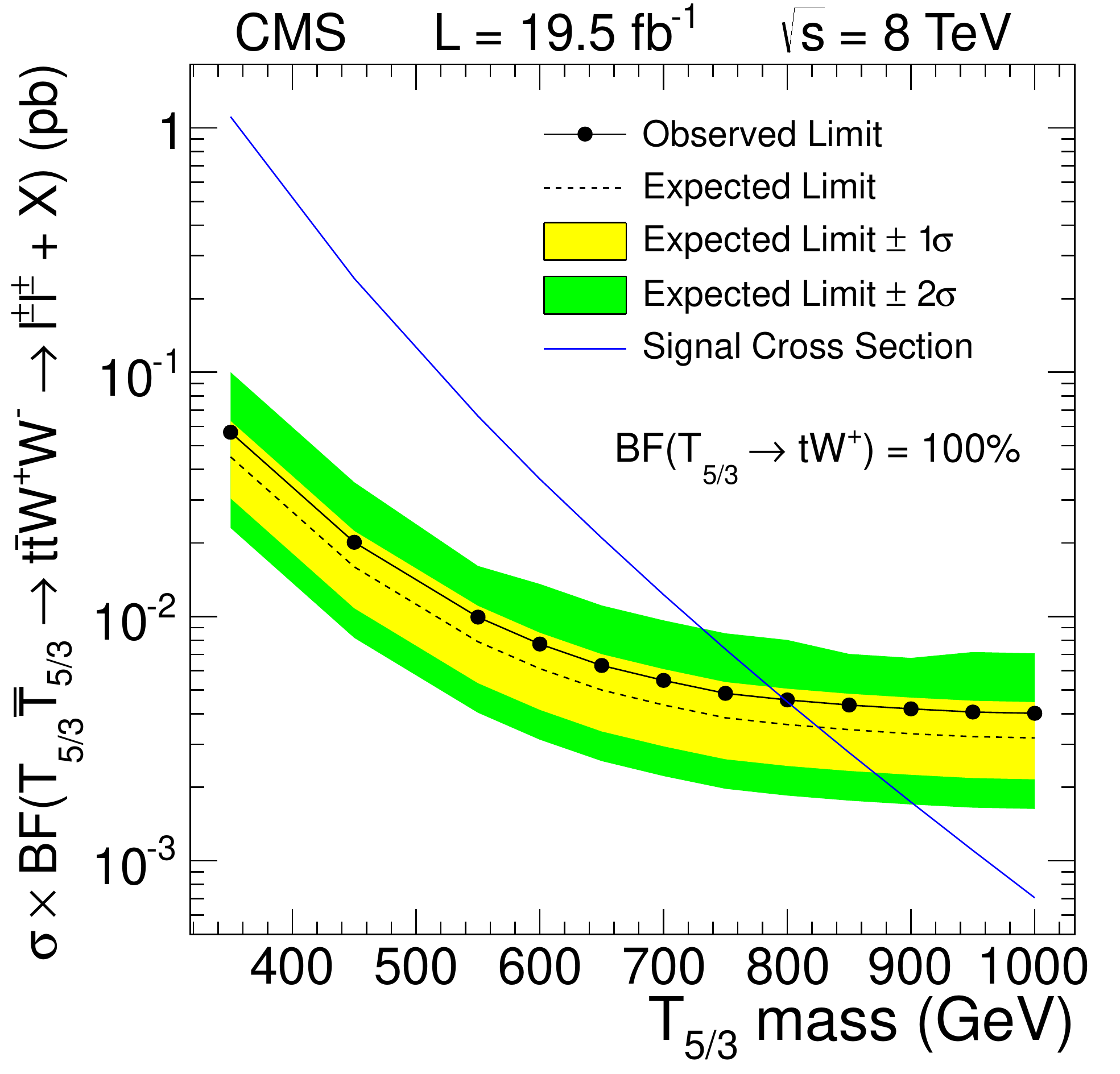} }
\caption{(a) Feynman diagram showing the production and decay of $\Tfivethree\Tfivethree \rightarrow W^+ t W^- t \rightarrow \ell^{\pm} \ell^{\pm} + \jets$. 
(b) Upper production cross section limits of \Tfivethree, assuming a decay into 2 same-sign charged leptons. 
 Figures from Ref.~\cite{Chatrchyan:2013wfa}. 
\label{fig:CMS-PAS-B2G-12-012}}
\end{figure}
The event selection requires two same-sign charged leptons ($e$ or $\mu$), \MET\ and jets. 
Top quarks are identified with the CMS top tagger~\cite{CMS:2014fya,Kaplan:2080ie} and boosted $W \rightarrow q \bar{q}'$ decays are also tagged with a jet substructure technique. 
The transition between boosted and resolved decays is solved elegantly by looking for a certain number of \emph{constituents}, $N_C \geq7$. 
Charged leptons or anti-\kt\ $R=0.5$ jets are counted as one constituent each, a $W$-tag is two constituents and a top-tag is three. 
\Tfivethree\ quarks with a mass lower than $m<800$~\GeV\ are excluded, $m<810$~\GeV\ exclusion expected, as illustrated in Fig.~\ref{F:CMS-PAS-B2G-12-012_Limit}. 

Another CMS search for top partners using the full 2012 dataset is the $\Tprime \Tprime \rightarrow Ht Ht \rightarrow 6b4q$ analysis~\cite{CMS:2014aka}. 
Here \Tprime\ is a vector-like quark with the same electric charge as the top quark. 
In this analysis, C-A \mbox{$R=1.5$} jets are studied. 
Top quarks are identified with the HEPTopTagger algorithm, and boosted Higgs bosons are identified by requiring that the C-A jet has at least two $b$-tagged sub-jets with an invariant mass \mbox{$m_{bb}>60$~\GeV}. 
The event signature is at least one top-tag and at least one Higgs-tag. 
The scalar sum of the transverse momenta of all selected objects, \HT, is shown in Fig.~\ref{F:CMS-PAS-14-002_TPrimeHistos_Main_HTSubJetsSingleHiggsTagBin}. The multi-jet (QCD) background is well suppressed.
\begin{figure}[htb]
\centering
\subfigure[\label{F:CMS-PAS-14-002_TPrimeHistos_Main_HTSubJetsSingleHiggsTagBin}]{
\includegraphics[width=0.39\textwidth]{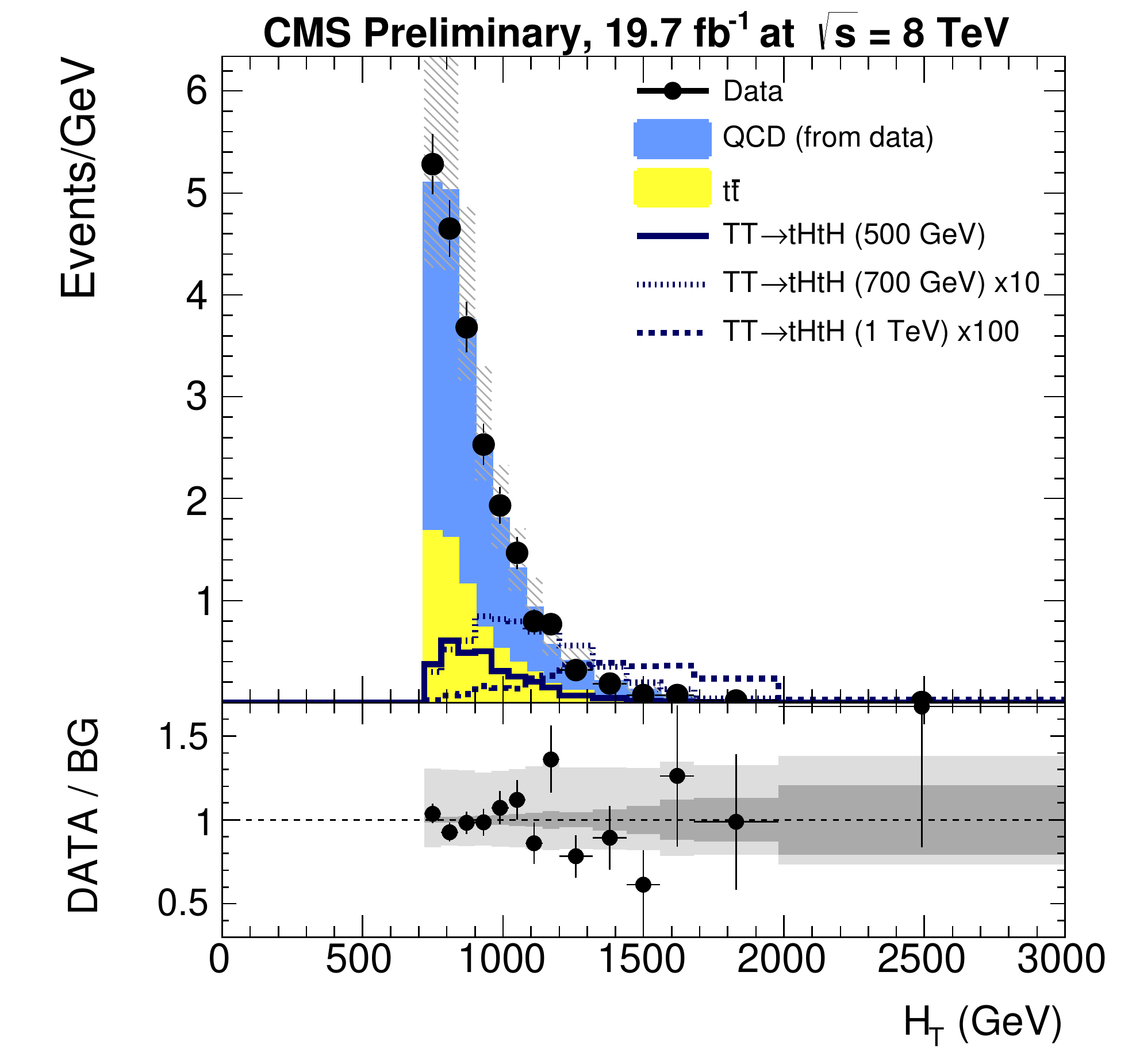} }
\subfigure[\label{F:CMS-PAS-14-002_limitplot_likelihood_bayesian}]{
\includegraphics[width=0.51\textwidth]{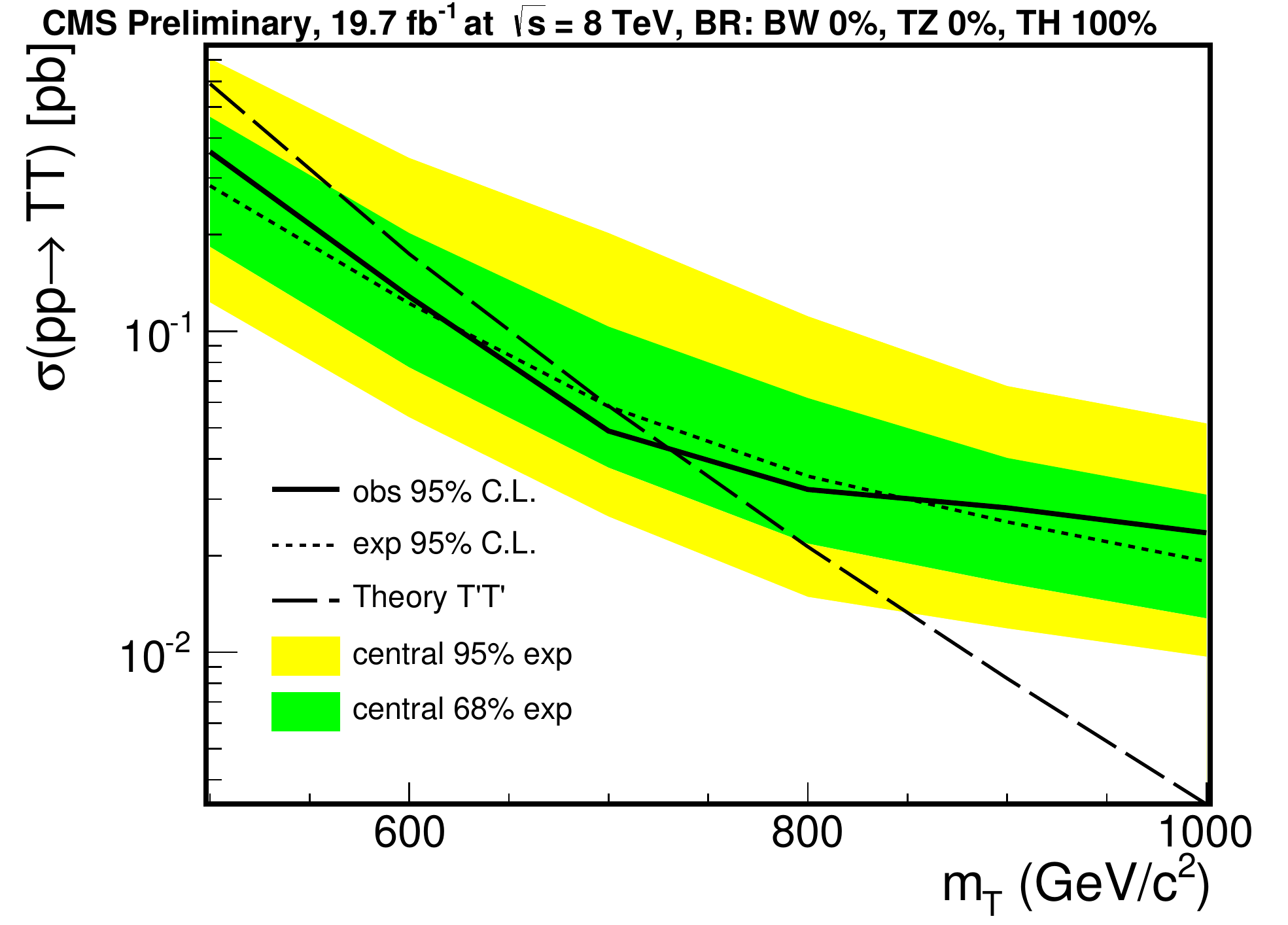} }
\caption{(a) The scalar sum of the transverse momenta of all selected objects, \HT. 
(b) Upper cross section limits of the \Tprime\ production, assuming $\Tprime \rightarrow Ht$. 
Figures from Ref.~\cite{CMS:2014aka}.
\label{fig:CMS-PAS-14-002_limitplot_likelihood_bayesian}}
\end{figure}
The analysis excludes $\Tprime$, assuming 100\% decay into $Ht$, for $m_{\Tprime} <747$~\GeV, ($m_{\Tprime}<701$~\GeV\ expected) as indicated in Fig.~\ref{F:CMS-PAS-14-002_limitplot_likelihood_bayesian}. 
Results with branching ratios including $Wb$ and $Zt$ are also given in the note~\cite{CMS:2014aka}. 

These limits are comparable with those from another CMS study, using the same dataset and model, but looking for final states including at least one lepton~\cite{Chatrchyan:2013uxa}, also employing top-tagging techniques. 
This clearly demonstrates that the usage of boosted top reconstruction techniques opens the final states with only hadrons for physics analyses at high transverse momenta, thus significantly increasing the amount of reconstructible top quark events for searches and measurements. 

\section{Summary and outlook}

Important physics results with boosted top quarks have been produced at the Tevatron and in Run-1 of the LHC. 
It is clear that the exclusion limits, and thereby the discovery potential, of new physical phenomena become stronger when employing substructure techniques. 
Plenty of \ttbar\ resonance searches have already been conducted, as well as the first top partner searches, e.g.~$\Tprime \rightarrow Ht$. 
A vast amount of searches involving top quarks could potentially benefit from including boosted techniques, since this increases the mass reach and enables searches in all-hadronic final states. 
Also boosted hadronic decays of bosons ($H$, $W$, $Z$) offer new possibilities to probe the SM at the highest accessible energies. 
With Run-2 of the LHC coming up, it is fair to assume that the boosted era has just begun.

\Acknowledgements
For fruitful discussions and useful comments when compiling my talk and the proceedings, I would like to thank 
L.~Asquith, R.~Brenner, G.~Chiarelli, K.~Cranmer, D.~Duda, R.~Erbacher, J.~Erdmann, J.~Ferrando, A.~Ferrari, S.~Fleischmann, L.~Gladilin, S.~Head, A.~Henrichs, A.~Jung, Y.~Peters, S.~Rappoccio, P.~Sinervo, S.~Strandberg, M.~Vos, H.~Wahl, C.~Wanotayaroj and J.~Zhong.

\end{document}

%% file: econfmacros.tex
\def\beq{\begin{equation}}
\def\eeq#1{\label{#1}\end{equation}}
\def\eeqn{\end{equation}}

\def\beqa{\begin{eqnarray}}
\def\eeqa#1{\label{#1}\end{eqnarray}}
\def\eeqan{\end{eqnarray}}

\let\bar=\overbar

\def\Dslash{\not{\hbox{\kern-4pt $D$}}}
\def\dslash{\not{\hbox{\kern-2pt $\del$}}}

\def\msb{{\bar{\ssstyle M \kern -1pt S}}}

\def\TeV{\ifmmode {\mathrm{\ Te\kern -0.1em V}}\else
                   \textrm{Te\kern -0.1em V}\fi}%
\def\GeV{\ifmmode {\mathrm{\ Ge\kern -0.1em V}}\else
                   \textrm{Ge\kern -0.1em V}\fi}%
\def\MeV{\ifmmode {\mathrm{\ Me\kern -0.1em V}}\else
                   \textrm{Me\kern -0.1em V}\fi}%
\def\keV{\ifmmode {\mathrm{\ ke\kern -0.1em V}}\else
                   \textrm{ke\kern -0.1em V}\fi}%
\def\eV{\ifmmode  {\mathrm{\ e\kern -0.1em V}}\else
                   \textrm{e\kern -0.1em V}\fi}%

\def\TeVc{\ifmmode {\mathrm{\ Te\kern -0.1em V}/c}\else
                   {\textrm{Te\kern -0.1em V}/$c$}\fi}%
\def\GeVc{\ifmmode {\mathrm{\ Ge\kern -0.1em V}/c}\else
                   {\textrm{Ge\kern -0.1em V}/$c$}\fi}%
\def\MeVc{\ifmmode {\mathrm{\ Me\kern -0.1em V}/c}\else
                   {\textrm{Me\kern -0.1em V}/$c$}\fi}%
\def\keVc{\ifmmode {\mathrm{\ ke\kern -0.1em V}/c}\else
                   {\textrm{ke\kern -0.1em V}/$c$}\fi}%
\def\eVc{\ifmmode  {\mathrm{\ e\kern -0.1em V}/c}\else
                   {\textrm{e\kern -0.1em V}/$c$}\fi}%

\def\TeVcc{\ifmmode {\mathrm{\ Te\kern -0.1em V}/c^2}\else
                   {\textrm{Te\kern -0.1em V}/$c^2$}\fi}%
\def\GeVcc{\ifmmode {\mathrm{\ Ge\kern -0.1em V}/c^2}\else
                   {\textrm{Ge\kern -0.1em V}/$c^2$}\fi}%
\def\MeVcc{\ifmmode {\mathrm{\ Me\kern -0.1em V}/c^2}\else
                   {\textrm{Me\kern -0.1em V}/$c^2$}\fi}%
\def\keVcc{\ifmmode {\mathrm{\ ke\kern -0.1em V}/c^2}\else
                   {\textrm{ke\kern -0.1em V}/$c^2$}\fi}%
\def\eVcc{\ifmmode  {\mathrm{\ e\kern -0.1em V}/c^2}\else
                   {\textrm{e\kern -0.1em V}/$c^2$}\fi}%

\def\ifb{fb$^{-1}$}

\def\sqrts{\ensuremath{\sqrt{s}}}

\def\Zprime{\ensuremath{Z^\prime}}
\def\Tprime{\ensuremath{T^\prime}}

\def\gKK{\ensuremath{g_{\mathrm{KK}}}}
\def\Tfivethree{\ensuremath{T_{5/3}}}

\def\antibar#1{\ensuremath{#1\bar{#1}}}

\def\ttbar{\antibar{t}}

\def\ppbar{\antibar{p}}

\def\ttbarljets{\ensuremath{\ttbar \rightarrow \ell + \mathrm{jets}}}
\def\ttbaralljets{\ensuremath{\ttbar \rightarrow \mathrm{jets}}}
\def\ljets{\ensuremath{\ell + \mathrm{jets}}}
\def\tlnub{\ensuremath{t \rightarrow \ell \nu b}}
\def\tqqb{\ensuremath{t \rightarrow q\bar{q}'b}}

\def\pt{\ensuremath{p_{\mathrm{T}}}}

\def\HT{\ensuremath{H_{\mathrm{T}}}} 
\def\MET{\ensuremath{E_{\mathrm{T}}^{\mathrm{miss}}}}

\def\mttbar{\ensuremath{m_{\ttbar}}} 
 
\def\mjet{\ensuremath{m_{\mathrm{jet}}}} 
\def\mjetone{\ensuremath{m_{\mathrm{jet1}}}} 
\def\mjettwo{\ensuremath{m_{\mathrm{jet2}}}} 
\def\pttop{\ensuremath{\pt^{\mathrm{top}}}} 
\def\ptjet{\ensuremath{\pt^{\mathrm{jet}}}} 

\def\dR{\ensuremath{\Delta R}} 
\def\kt{\ensuremath{k_t}} 
\def\jet{\ensuremath{\mathrm{jet}}}
\def\jets{\ensuremath{\mathrm{jets}}}

\newcommand{\kommentar}[1]{}
 % ...
\kommentar{%
   the text in the brackets is interpreted as comment...
  }%